\documentclass[12pt]{iopart}

\usepackage{iopams}

\usepackage{graphicx}

\newtheorem{theorem}{Theorem}
\newtheorem{lemma}{Lemma}
\newtheorem{corollary}[theorem]{Corollary}

\newcommand{\sn}{\textrm{sn}}
\newcommand{\cn}{\textrm{cn}}

\begin{document}
\title[Three-Body Choreographies in Given Curves]
{Three-Body Choreographies in Given Curves}
\author{
Hiroshi~Ozaki$^1$,
Hiroshi~Fukuda$^2$ and
Toshiaki~Fujiwara$^3$
}
\address{$^1$General Education Program Center, Tokai University,
317 Nishino, Numazu, Shizuoka 410-0395, Japan}

\address{$^{2,3}$College of Liberal Arts and Sciences, Kitasato University, 
1-15-1 Kitasato, Sagamihara, Kanagawa 228-8555, Japan}%

\eads{
	\mailto{$^1$ozaki@keyaki.cc.u-tokai.ac.jp}, 
	\mailto{$^2$fukuda@kitasato-u.ac.jp},
	\mailto{$^3$fujiwara@kitasato-u.ac.jp}
}

\begin{abstract}
As shown by Johannes Kepler in 1609,
in the two-body problem,
the shape of the orbit, a given ellipse,
and a given non-vanishing constant angular momentum
determines the motion of the planet 
completely.

Even in the three-body problem, in some cases,
the shape of the orbit, conservation of the centre of mass
and a constant of motion (the angular momentum or the total energy)
determines the motion of the 
three bodies.
We show, by a geometrical method, that
choreographic motions,
in which equal mass three bodies chase each other 
around 
a same curve,
will be uniquely determined for the following two cases.
(i) Convex curves that have  point symmetry 
and non-vanishing angular momentum are given.
(ii) Eight-shaped curves which are similar to the curve for the figure-eight solution
and the energy constant are given.

The reality of the motion should be tested
whether the motion satisfies an equation of motion or not.

Extensions of the method  for generic curves are shown.
The extended methods are applicable to generic curves which does  not have point symmetry.
Each body may have its own curve and its own non-vanishing masses.
   
\end{abstract}


\section{Introduction}
\label{sec:Intro}
The three-body figure-eight solution is
 one of the solutions of the planar equal mass three-body problem
 %
under the Newtonian gravity.
In this solution, three bodies chase each other around a fixed
eight-shaped curve.
It was found numerically by Moore~\cite{moore}
and its existence was proved by Chenciner and Montgomery~\cite{chenAndMont}.

Only a few is known about the eight-shaped curve.
Sim\'{o} showed numerically that the curve 
cannot
be 
expressed by
algebraic curves
of order $4, 6, 8$~\cite{simo1, simo2}.
Chenciner and Montgomery~\cite{chenAndMont} showed that
the curve is a ``star-shape'', namely, a ray from the origin 
meets the curve at most once.
%
Fujiwara and Montgomery~\cite{fujiwaraMontgomery} proved that
the each lobe of the eight-shaped curve 
is convex.

On the other hand,
the present authors found a parameterization
$q_i(t)=\big( x_i(t), y_i(t) \big)$
of the lemniscate of Bernoulli $(x^2+y^2)^2=x^2-y^2$,
which 
satisfies
 an equation of motion
under an inhomogeneous potential~\cite{ffoLem},
\begin{equation}
\frac{d^2 q_i}{dt^2} = - \frac{\partial V}{\partial q_i},
\end{equation}
\begin{equation}
\label{eq:thePotentialForLemniscate}
V=\sum_{i<j}\left( \frac{1}{2} \ln r_{ij} -\frac{\sqrt{3}}{24} r_{ij}^2 \right),
\end{equation}
where $r_{ij}=|q_i-q_j|$ 
is the mutual distance between the body $i$ and $j$.

An interesting point of their approach is that
they started the arguments from the lemniscate curve,
without any assumption for the potential.
They showed that there 
is a parameterization $q_i(t)$ of the curve
that 
keeps the geometric centre of mass being at the origin $\sum_i q_i(t)=0$
and 
keeps the angular momentum being zero $\sum q_i(t) \times dq_i(t)/dt=0$ for all $t$.
Using this parameterization, 
they searched for what kind of potential
can support this motion.
Finally, they found the potential (\ref{eq:thePotentialForLemniscate}).

Then, a question arises.
Does a similar approach work for other eight-shaped curves?
Namely, can we determine the three-body motion $q_i(t)$
if the shape of the orbit for the figure-eight solution is known?

This approach works for the two-body problem
as shown by Johannes Kepler in ``Astronomia Nova'' published in 1609.
In 
his 
book, he stated the first law, 
planets move in elliptical curve with the sun at one focus.
Then, his second law, now 
which
 is known as the conservation of the angular momentum,
determines the motion of a planet in the ellipse
if we give a non-vanishing constant angular momentum.

In this paper, we 
show that this approach works in the three-body  
problem.
Namely, for some curves,
conservation of the centre of mass and a constant of the motion
(the angular momentum or the total energy) determine the three-body motion.
Here, in the two-body problem,
the total energy constant, instead of vanishing angular momentum,
determines the motion if the orbit is linear.

To show the main idea, let us observe how the
condition for
geometrical centre of mass being at the origin,
\begin{equation}
\label{eq:CenterOfMass}
q_1+q_2+q_3=0,
\end{equation}
determines the mutual positions of the three bodies.
%
Let us consider a unit circle $|z|=1$ in the complex plane
and a position $q_3=\exp(i \phi)$ on the circle.
We know that
the set of two points $\{q_1, q_2\}$ on the same circle
that satisfy $q_1+q_2+q_3=0$ is $\{q_1,q_2\}=\{\exp(i \phi +2\pi/3),\exp(i \phi -2\pi/3)\}$.
%
On the other hand, it is obvious that the two points are the cross points
of the original circle $|z|=1$ and the unit circle $|z+q_3|=1$,
which is the parallel translation $z \mapsto z-q_3$ of the original circle.

One of the authors, Ozaki, noticed that
this is not an accident.
He found the following theorem.
\begin{theorem}[Construction of three points]
\label{Ozaki}
If a curve $\gamma$ in $\mathbb{R}^d$ with 
$d =2,3,4, \dots$ 
is invariant under the inversion
$q \mapsto -q$
then the set 
$\{ \{q_1, q_2\} | q_1, q_2 \in \gamma, q_1+q_2+q_3=0\}$
for a given $q_3 \in \gamma$ 
is equal to the set
$\{ \{q, q^*\} | q\in \gamma \cap \gamma_{\|}\}$
where $\gamma_{\|}$ is 
the parallel translation
$q \mapsto q-q_3$
of the curve $\gamma$ and $q^*=-q-q_3$.
\end{theorem}
%
%
This theorem gives us a method to find three points
$q_1, q_2, q_3$ with $q_1+q_2+q_3=0$
on a point symmetric curve with respect to the origin.
This theorem states that for such curve
and for a given $q_3 \in \gamma$,
(i) if there is a pair $q_1,q_2\in \gamma$ that satisfy $q_1+q_2+q_3=0$
then the points $q_1$ and $q_2$ should be the cross points of $\gamma$ and $\gamma_{\|}$,
and inversely,
(ii) if 
a cross point 
$q$
of $\gamma$ and $\gamma_{\|}$ exists
then the point 
$q^*=-q-q_3$ 
is also a cross point of the same curves,
and 
%
$q + q^* + q_3 = 0$
is satisfied.

See Figures \ref{fig:figConvexCurve}, 
\ref{fig:beforeCritical}, \ref{fig:critical} and \ref{fig:afterCritical}.
Figure \ref{fig:figConvexCurve} shows the situation for convex curve
that is invariant under $q \mapsto -q$.
This figure suggests that
the pair of the cross points $\{q_1,q_2\}=\{q, q^* \}$, 
namely the solution of $q_1+q_2+q_3=0$, is unique for $q_3 \in \gamma$.
%
%
Figures \ref{fig:beforeCritical}, \ref{fig:critical} and \ref{fig:afterCritical} 
show the situation for an eight-shaped curve.
These figures suggest that 
there are two pairs of the cross points, 
trivial pair 
$\{O,-q_3\}$ 
in which three points $-q_3$, $O$, and $q_3$ are collinear, 
and one non-trivial pair $\{q_1,q_2\}$.

For these cases, we can show that
the (non-trivial) pair $\{q_1,q_2\}$ are determined uniquely
for a given $q_3\in \gamma$.
%
Moreover, we can show that 
if we move $q_3$ 
around the whole curve,
the points $q_1$ and $q_2$ move smoothly 
and strongly monotonically
around the whole curve
without collisions.
Thus the motion around such curve is determined uniquely modulo
time re-parameterization,
$q_i(t) \mapsto q_i(\tau(t))$ with 
a function $\tau(t)$.

In Section \ref{sec:Extension},
proofs of Theorem \ref{Ozaki} are given.
In Theorem \ref{Ozaki},
the points $q_1$ and $q_2$ are assumed to be constrained on a same curve,
the curve are assumed to be point symmetric,
and 
the
three bodies are assumed to have the same mass.
We can remove there assumptions.
Extensions of Theorem \ref{Ozaki} are also given in Section  \ref{sec:Extension}.
%
In Section \ref{sec:convexCurve},
considering the geometrical property of the cross points of
the convex curve and its translation we prove the uniqueness 
and the smoothness of $\{q_1,q_2\}$ 
for a given $q_3$.
Then, 
we show that motions of equal mass three bodies
in planar point symmetric convex curves with respect to 
%
the origin
are uniquely determined if non-vanishing angular momentum are given.
In 
Section \ref{sec:Fig8}, 
we show, the main result, 
the uniqueness of the motions of equal mass three bodies
in planar eight-shaped curves
%
if the energy constant are given.
Section \ref{sec:summary} is the summary and discussions.

\section{Constructions of three points}
\label{sec:Extension}
In this section,
we prove  some geometrical constructions of three points
whose geometrical centre of mass is fixed to the origin,
namely Theorem~\ref{Ozaki} and its extensions.

To prove Theorem \ref{Ozaki},
it is convenient to introduce a function $\Gamma$ of $q \in \mathbb{R}^d$
that characterizes the curve $\gamma$,
such that,
\begin{equation}
\Gamma(q)=0 \iff q \in \gamma.
\end{equation} 
Then, the invariance under the inversion $q \to -q$ of the curve $\gamma$
is expressed by
\begin{equation}
\label{Sym}
\Gamma(q)=0 \iff \Gamma(-q)=0.
\end{equation}

\noindent
\textbf{Proof of Theorem \ref{Ozaki}:}
If there is a solution $q_1, q_2\in \gamma$ with $q_1+q_2+q_3=0$,
these points satisfy
\begin{equation}
\Gamma(q_1)=\Gamma(q_2)=0.\label{eq:q}
\end{equation}
Using $q_1+q_2+q_3=0$ and the symmetry (\ref{Sym}), we get
\begin{eqnarray}
\Gamma(q_1+q_3)=\Gamma(-q_2)=\Gamma(q_2)=0,\label{eq:q1Plusp}\\
\Gamma(q_2+q_3)=\Gamma(-q_1)=\Gamma(q_1)=0.\label{eq:q2Plusp}
\end{eqnarray}
The equations (\ref{eq:q})
--
(\ref{eq:q2Plusp})
show that the points $q_1$ and $q_2$ satisfy the equations
\begin{equation}
\Gamma(q)=0 \mbox{ and } \Gamma(q+q_3)=0.
\label{eq:GammaGamma}
\end{equation}
Namely, $q_1, q_2 \in \gamma \cap \gamma_{\|}$.

Inversely, if there is a cross point $q$ of the curves $\gamma$ and $\gamma_{\|}$,
$q$ satisfies
the equation (\ref{eq:GammaGamma}).
Then, $q\in \gamma$ by $\Gamma(q)=0$, and
$(-q-q_3) \in \gamma$ by $\Gamma(-q-q_3)=\Gamma(q+q_3)=0$.
Moreover,
$q+(-q-q_3)+q_3=0$ is satisfied.

\noindent
\textbf{Remark for Theorem \ref{Ozaki}:}
In the above proof,
the condition $q_3 \in \gamma$ is not used.
Actually, Theorem \ref{Ozaki} is true for $q_3 \in \mathbb{R}^d$.

Now, let us state an extension of Theorem \ref{Ozaki}.

\begin{theorem}
\label{Ozaki-Fukuda}
For a given set 
$\gamma_1, \gamma_2 \subset \mathbb{R}^d$ and
$q_3 \in \mathbb{R}^d$,
we have the following equalities,
\begin{eqnarray}
\{ \{q_1, q_2\} | q_1\in \gamma_1, q_2\in \gamma_2, q_1+q_2+q_3=0\}
&=&\{ \{q_1, q_1^*\} | q_1 \in \gamma_1 \cap \gamma_2^* \}\label{eq:gen1}\\
&=&\{ \{q_2, q_2^*\} | q_2 \in \gamma_1^* \cap \gamma_2 \}\label{eq:gen2},
\end{eqnarray}
where ${}^*$ represents a map
$q \mapsto q^*=-q-q_3$
and $\gamma^*$ is the image of $\gamma$ by this map.
\end{theorem}

\noindent
\textbf{Proof of Theorem \ref{Ozaki-Fukuda}:}
We prove the equation (\ref{eq:gen1}).
The 
proof for the equation (\ref{eq:gen2}) is 
similar.
If $q_1$ and $q_2$ satisfy
$q_1\in \gamma_1$, $q_2\in \gamma_2$ and $q_1+q_2+q_3=0$,
then $q_1=-q_2-q_3=q_2^* \in \gamma_2^*$.
Therefore, $q_1 \in \gamma_1 \cap \gamma_2^*$ and $q_2=q_1^*$.
Inversely,
if $q_1$ and $q_2$ are given by
$q_1 \in \gamma_1 \cap \gamma_2^*$ and $q_2=q_1^*$,
then
$q_1 \in \gamma_1 \cap \gamma_2^* \subset \gamma_1$,
$q_2 = q_1^* \in \gamma_1^* \cap \gamma_2 \subset \gamma_2$.
Moreover,
by the definition of $q_2=q_1^*$ we get $q_1+q_2+q_3=0$.

If $q_1$, $q_2$ and $q_3$ move around the same set $\gamma$,
we have the following corollary by simply making $\gamma_1=\gamma_2=\gamma$.

\begin{corollary}
\label{cor:OzFkd}
For a given set 
$\gamma \subset \mathbb{R}^d$ and
$q_3 \in \gamma$,
we have the following equality,
\begin{equation}
\{ \{q_1, q_2\} | q_1, q_2\in \gamma, q_1+q_2+q_3=0\}
=\{ \{q, q^*\} | q \in \gamma \cap \gamma^* \}\label{eq:gen3}.
\end{equation}
\end{corollary}

We do not assume any symmetry for the set $\gamma$ in this corollary.
So, this corollary can be used to make 
equal mass three-body motions  
in given curves
with no symmetry.

\begin{figure}[h] 
   \centering
   \includegraphics[width=5cm]{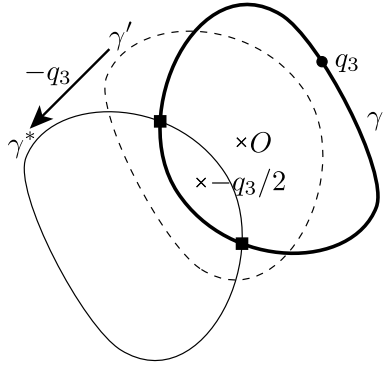}
      \caption{The curve $\gamma^*$ that is the image of $\gamma$
   by the map $q \mapsto q^*=-q-q_3$ can be drawn by two ways.
   i) Make the inversion of $\gamma$ with respect to the point $-q_3/2$.
   ii) Draw the inversion $q \mapsto -q$ of $\gamma$
   with respect to the origin, to get $\gamma'$.
  Then make parallel translation
   $q \mapsto q-q_3$ of $\gamma'$.
   }
   \label{fig:genericConvex}
\end{figure}

Note that the map $q \mapsto q^*=-q-q_3$ can be decomposed into
the map $q \mapsto -q$ followed by the map $q \mapsto q-q_3$.
Therefore, the curve $\gamma^*$ can be made  by
the two steps. First make inversion $\gamma'$ of $\gamma$ by $q \mapsto -q$,
then make parallel translation
$\gamma'_{\|}$ of $\gamma'$ by $q \mapsto q-q_3$.
See Figure \ref{fig:genericConvex}.
For the case $\gamma$ is invariant under the inversion, $q \mapsto -q$,
then $\gamma'=\gamma$ and
$\gamma^*=\gamma_{\|}$.
Thus, we get another proof of Theorem \ref{Ozaki}.

\textbf{Remark for Theorem \ref{Ozaki-Fukuda}:} 
For three bodies with general masses $m_i \ne 0$,
the centre of mass being at the origin is defined by
\begin{equation}
\label{eq:mass}
\sum_{i=1,2,3} m_i q_i=0.
\end{equation}
For this case,
let 
\begin{equation}
\label{eq:massScaling}
\tilde{q}_i=m_i q_i
\end{equation}
and $\tilde{\gamma}_i$ be image of the curves $\gamma_i$ by the map
$q_i \mapsto \tilde{q_i}=m_i q_i$.
Then the conditions
$q_1\in \gamma_1$, $q_2 \in \gamma_2$ and 
Eq.~(\ref{eq:mass})
are equivalent to 
$\tilde{q}_1 \in \tilde{\gamma}_1$, $\tilde{q}_2 \in \tilde{\gamma}_2$
and 
\begin{equation}
\sum_{i=1,2,3}\tilde{q}_i=0.
\end{equation}
Then, we can apply Theorem \ref{Ozaki-Fukuda}
for $\tilde{q}_i$ and $\tilde{\gamma}_i$.
Once we find the positions $\tilde{q}_1$ and $\tilde{q}_2$,
we get the positions $q_i = m_i^{-1}\tilde{q}_i$ for $i=1,2$.

\section{Three-body choreography in point symmetric convex curve}
\label{sec:convexCurve}
\subsection{Motion in point symmetric convex curve}
\label{subsec:convexCurve}
In this subsection,
as a simple application of Theorem~\ref{Ozaki},
we construct a 
equal mass 
three-body 
motion in a given closed convex curve $\gamma$
that is invariant under the inversion $q \mapsto -q$.
We 
assume that the curvature  
is not zero
everywhere on $\gamma$.

\begin{figure}[h]
   \centering
   \includegraphics[width=5cm]{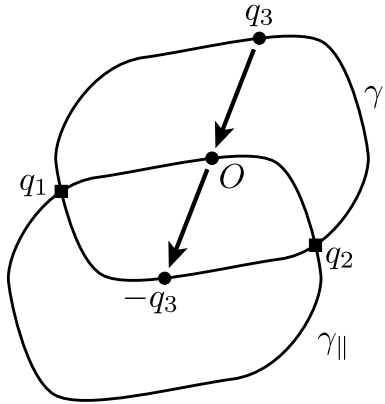} 
   \caption{The closed convex curve $\gamma$ is 
   point symmetric with respect to the origin $O$.
    For a given 
   point $q_3 \in \gamma$,
   the two points on $\gamma$ that satisfy $q_1+q_2+q_3=0$ are
   given by the cross points of $\gamma$ and $\gamma_{\|}$.}
   \label{fig:figConvexCurve}
\end{figure}

\begin{theorem}
\label{th:convex}
If closed 
planar
convex curve $\gamma$ is 
invariant under the inversion $q \mapsto -q$ 
and its curvature    
is not zero
everywhere on $\gamma$,
the solutions of $q_1+q_2+q_3=0$ with $q_1, q_2 \in \gamma$ 
for a given $q_3 
\in \gamma$ are unique.
Moreover, when $q_3$ moves around $\gamma$,
the motion $q_i(\sigma)$, $i=1,2$ are smooth, i.e., $|dq_i/d\sigma|<\infty$, 
and strongly monotonic, i.e., $dq_i/d\sigma\ne 0$, 
where $\sigma$ is the curve length for $q_3$.
\end{theorem}
For a given
 $q_3 \in \gamma$, the pair of positions $\{q_1,q_2\}$
is given by Theorem~\ref{Ozaki}
as $\{q_1, q_2\}=\{ q, q^*\}$.
%
First, we show the uniqueness of the pair $\{q_1,q_2\}$.
%
As shown in Figure \ref{fig:figConvexCurve}, 
the map $q \mapsto q-q_3$
maps
$q_3 \in \gamma$ to the origin $O$,
$O$ to the point $-q_3 \in \gamma-\gamma_\|$,
and 
$-q_3$ to the $-2q_3 \in \gamma_{\|}-\gamma$.
Therefore, the curve $\gamma_{\|}$ starts at the origin $O$
which is surrounded by $\gamma$ and 
passes the point $-2q_3$ which is outside of $\gamma$.
Then, there are at least two points in $\gamma \cap \gamma_{\|}$.
On the other hand, 
$\gamma \cap \gamma_{\|}$
has at most two elements by  Lemma
\ref{lemma:convexcurve} in \ref{sec:convexcurve}.
%
Therefore, $\gamma \cap \gamma_{\|}$ has exactly two elements.
 Thus we find an unique pair
$\{q_1, q_2 \}=\{q, q^* \}$ that satisfy $q_1+q_2+q_3=0$
by Theorem~\ref{Ozaki}.
%

Let us move the point $q_3$ 
around the whole curve 
to one direction,
namely using the curve length $\sigma$ for $q_3$,
\begin{equation}
q_3=q_3(\sigma)
\mbox{ with } 
\left|\frac{dq_3}{d\sigma}\right|=1.
\label{eq:unit_speed}
\end{equation}
Then $q_1$ and $q_2$ are uniquely parameterized by the same 
parameter $\sigma$.
To 
prove that $q_i(\sigma)$ for $i=1,2$ 
are smooth and strongly monotonic functions of $\sigma$, i.e.,
\begin{equation}
\left| \frac{dq_i(\sigma)}{d\sigma} \right| < \infty \mbox{ and } 
\frac{dq_i(\sigma)}{d\sigma} \ne 0,
\end{equation}
%
note that 
when $q_3$ moves around $\gamma$ with some speed, 
$\gamma_\|$ moves to the opposite direction with the same speed 
since the center of $\gamma_\|$ is $-q_3$.
Therefore,
\begin{lemma}
In Theorem \ref{Ozaki},
(i) if the tangent lines to the curves $\gamma$ and $\gamma_{\|}$ at $q$ are distinct,
then $q(\sigma)$ and $q^*(\sigma)$
are smooth functions of
$\sigma$ where $\sigma$ is the curve length for $q_	3$.
(ii) Further, if the tangent lines to the curve $\gamma$ at $-q_3$ is not parallel to
the tangent line to the curve $\gamma_{\|}$ at $q$,
then $q(\sigma)$ and $q^*(\sigma)$ is smooth and strongly monotonic function of 
$\sigma$.
\label{lem:smooth}
\end{lemma}
\begin{figure}[h]
   \centering
   \includegraphics[width=7cm]{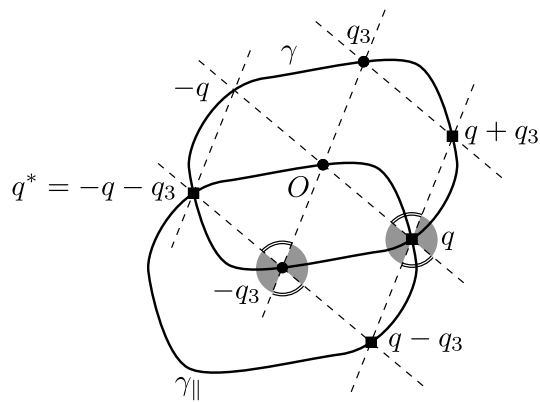} 
   \caption{Lemma \ref{lem:smooth} for closed convex curve with point symmetry.
We denote one of the cross points of $\gamma$ and $\gamma_\|$ by $q$. 
(i) The parallelogram $\alpha q + \beta q_3$, $0 \le \alpha, \beta \le 1$ 
   is included inside $\gamma$, 
   thus, at $q$, the tangent line to the curve $\gamma$ passes in the shaded area,
   while
   the tangent line to the curve $\gamma_\|$ at $q$ passes in the non shaded area
   because
   the parallelogram $\alpha q - \beta q_3$, $0 \le \alpha, \beta \le 1$ 
   is included inside the $\gamma_{\|}$.
   Therefore,
   the tangent lines at $q$ to the lines $\gamma$ and $\gamma_{\|}$
   are distinct.
(ii)  The parallelogram $-\alpha q - \beta q_3$, $0 \le \alpha, \beta \le 1$ 
   is included inside $\gamma$,
   thus the tangent line to the curve $\gamma$ at $-q_3$ passes in the shaded area.
   Therefore,
   the tangent line to $\gamma_{\|}$ at $q$
   and the tangent line to $\gamma$ at $-q_3$
   are not parallel. 
   }
   \label{fig:smooth_convex}
\end{figure}
For $\gamma$ in Theorem \ref{th:convex}, from Figure \ref{fig:smooth_convex}
it is clear that both the conditions (i) and (ii) are 
satisfied for all $q_3 \in \gamma$,
therefore $q_1(\sigma)$ and $q_2(\sigma)$ are smooth and strongly monotonic by 
Lemma \ref{lem:smooth}. 
Then Theorem \ref{th:convex} is proved.

Now, we demand the angular momentum 
\begin{equation}
\label{eq:c}
c
=\sum_{i=1,2,3} q_i(\sigma(t)) \times \frac{dq_i(\sigma(t))}{dt}
=\frac{d\sigma}{dt}
	\sum_{i=1,2,3} q_i(\sigma) \times \frac{dq_i(\sigma)}{d\sigma}
\end{equation}
to be constant
in order to investigate the motion in $\gamma$.
For the convex curve, we have
\begin{equation}
J(\sigma)=
\sum_{i=1,2,3} q_i(\sigma) \times \frac{dq_i(\sigma)}{d\sigma}
\ne 0.
\end{equation}
This is because,
%
if $J(\sigma)=0$,
three tangent lines must meet at a point
by the three tangents theorem
found by the present authors \cite{ffoLem,ffkoy,fujiwaraMontgomery},
whereas there are at most two tangent lines 
to the convex curve from one point.
Thus, the time dependence of $\sigma$ is determined by
\begin{equation}
\frac{d\sigma}{dt}
=\frac{c}{J(\sigma)}.
\end{equation}
Accordingly
the motion $q_i(\sigma(t))$ is determined uniquely
by $c$.

Since $J(\sigma) \ne 0$,
the sign of $d\sigma/dt$ is fixed.
Let us take the sign of $c$ be positive,
then the points $q_i(\sigma(t))$ move anti-clockwise.
%
By the uniqueness of $\{q_1, q_2\}$
if $q_3$ moves around the whole $\gamma$,
the other points $q_1$ and $q_2$ 
move around the  whole curve
without collision.
So, we can name the three points $q_1$, $q_2$ and $q_3$ in anti-clockwise order.
%
%

In the following, we write $q_i(t)=q_i(\sigma(t))$ for simplicity.
At time $t=0$, the three  points are at $q_i(0)$.
As time passing,
the point $q_1(t)$ moves around the curve toward the point $q_2(0)$,
and at some time $t_0$, $q_1(t)$ reaches to the point $q_2(0)$.
Then, we have
\begin{equation}
q_1(t_0)=q_2(0),
q_2(t_0)=q_3(0),
q_3(t_0)=q_1(0),
\end{equation}
because one position determines the other two positions uniquely.
Then, the motion for $t_0 \le t \le 2t_0$ is also determined by
the motion $q_i(t)$ for $0\le t \le t_0$ as follows,
\begin{equation}
q_1(t) = q_2(t-t_0),
q_2(t) = q_3(t-t_0),
q_3(t) = q_1(t-t_0),
\end{equation}
because again one position determines the other two positions uniquely
and the name shift $1 \to 2$, $2 \to 3$ and $3 \to 1$ is equivalent to
the time shift $t \to t+t_0$.

We can proceed the same step over and over again, 
therefore we have a periodic motion
of $q_i(t)$ with period $T=3t_0$ and the motion is described by
\begin{eqnarray}
q_2(t)=q_1(t+T/3),\\
q_3(t)=q_1(t+2T/3).
\end{eqnarray}
We would like to call this periodic motion with an equal time spacing 
a ``choreography" in the point symmetric convex curve.
We should note that
there is no guarantee for this motion
to satisfy some equation of motion.

\subsection{Three-body choreography in an ellipse}
In this subsection, let us assume the curve $\gamma$ is an ellipse
\begin{equation}
\label{curveOval}
\frac{x^2}{a^2}+\frac{y^2}{b^2}=1
\end{equation}
with constants $a, b>0$.
This is convex and invariant under the inversion with respect to the origin,
$(x,y) \mapsto (-x,-y)$.
Therefore, 
by Theorem \ref{th:convex},
three-body choreography in this curve
satisfying
$q_1+q_2+q_3=0$ 
with constant angular momentum $c$
is determined uniquely.
For this case,
we can construct a choreography explicitly
and show that this choreography satisfies the equation of motion
for  harmonic oscillators.

This ellipse is parameterized by
\begin{equation}
q(\tau)=\left(x(\tau),y(\tau)\right)=\left( a\cos(\tau), b\sin(\tau)\right)
\end{equation}
with an arbitrary 
parameter $\tau$.
Then, the points
\begin{eqnarray}
q_1(\tau)=q(\tau),
q_2(\tau)=q(\tau+2\pi/3),
q_3(\tau)=q(\tau+4\pi/3)\label{eq:q1To3}
\end{eqnarray}
satisfy
$q_1+q_2+q_3=0$.
Since,
\begin{equation}
q(t)\times \frac{dq(\tau)}{d\tau}=ab,
\end{equation}
we get
\begin{equation}
\sum q_i(\tau)\times \frac{dq_i(\tau)}{d\tau}=3ab.
\end{equation}
Therefore,
the equation
$c=d\tau/dt \sum q_i(\tau)\times dq_i(\tau)/d\tau$
determines
\begin{equation}
\frac{d\tau}{dt}=\frac{c}{3ab}.
\end{equation}
%
Thus, the three-body choreography in the ellipse is uniquely determined by
\begin{equation}
q(t)
=\left( a \cos( \omega t), b \sin(\omega t) \right)
\mbox{ with } \omega = \frac{c}{3ab}
\end{equation}
and 
\begin{eqnarray}
q_1(t)=q(t),
q_2(t)=q\left(t+\frac{2\pi}{3\omega}\right),
q_3(t)=q\left(t+\frac{4\pi}{3\omega}\right)
\label{eq:t}
\end{eqnarray}

Obviously, $q_i(t)$ satisfies the equation of motion for the harmonic oscillator
\begin{equation}
\frac{d^2 q_i(t)}{dt^2} = -\omega^2 q_i(t).
\end{equation}
Using $q_1+q_2+q_3=0$,
we get $q_i=\sum_{j\ne i} (q_i-q_j)/3$.
Therefore, $q_i$ satisfy the following equations of motion,
\begin{eqnarray}
\frac{d^2q_i}{dt^2}
=\frac{\omega^2}{3} \sum_{j\ne i} (q_j-q_i)
=-\frac{\partial V}{\partial q_i},
\end{eqnarray}
with
\begin{equation}
V=\frac{\omega^2}{6}
	\sum_{i\ne j} |q_i-q_j|^2.
\end{equation}
Therefore,
the choreography in an ellipse is realized by the Hamiltonian
\begin{equation}
H=\sum_i \frac{|p_i|^2}{2} + \frac{\omega^2}{6} \sum_{i\ne j} |q_i-q_j|^2.
\end{equation}
For this choreography,
the kinetic energy $K$, the potential energy $V$
and the moment of inertia $I$ are the following constants,
\begin{eqnarray}
K
=\frac{1}{2}\sum_i \left| \frac{dq_i}{dt} \right|^2
=\frac{3\omega^2}{4}(a^2+b^2),\\
V
=\frac{\omega^2}{6}
	\sum_{i\ne j} |q_i-q_j|^2
=\frac{3\omega^2}{4}(a^2+b^2),\\
I
=\sum_i \left| q_i \right|^2
=\frac{3}{2}(a^2+b^2).
\end{eqnarray}

\section{Three-body choreography in an eight-shaped curve}
\label{sec:Fig8}
We consider the eight-shaped curve $\gamma$ 
defined by the following properties.
(I) The $\gamma$ is invariant under the inversion $x \mapsto -x$ or $y \mapsto -y$.
(II) The three points $O= (0,0)$ and $ (\pm 1,0)$ are on the $\gamma$.
(III) In the first quadrant, the $\gamma$ is described by a function as
$(x, f(x))$ for $0 \le x \le 1$
that satisfies $f(0)=f(1)=0$ and $f(x)>0$ for $0 < x < 1$.
(IV) For the smoothness of the curve, 
\begin{equation}
f'(0)=\lim_{x \to +0}f'(x) >0
\mbox{ and }\lim_{x \to 1-0}f'(x) \to -\infty.
\end{equation}
These properties (I)--(IV) are acceptable as those for the usual eight-shaped curves. 

We look for the 
solution $q_1, q_2 \in \gamma$
satisfying 
$q_1+q_2+q_3=0$ 
 for a given $q_3 \in \gamma$ 
in  
the cross points $\gamma \cap \gamma_\|$ according to Theorem \ref{Ozaki}.
Since 
the origin $O \in \gamma$,
we find the trivial solution $\{q_1,q_2\}=\{O,-q_3\}$ for $q_3 \in \gamma$
in which the three points $q_3$, $O$ and $-q_3$ are collinear.
See the Figures \ref{fig:beforeCritical}, \ref{fig:critical} and \ref{fig:afterCritical}.
This trivial solution, however, has no physical importance
since it does not conserve the angular momentum
\begin{equation}
\sum_{i=1,2,3} q_i \times \frac{dq_i}{dt}
= 0 \times 0 + (-q_3)\times \frac{d(-q_3)}{dt}+q_3\times \frac{dq_3}{dt}
=2 q_3 \times \frac{dq_3}{dt},
\end{equation}
which changes the sign at $q_3=0$.
Moreover, 
these three points will go to the three-body collision at the origin
when $q_3 \to O$.

On the other hand,
Figures \ref{fig:beforeCritical} and \ref{fig:afterCritical} suggest that
there is just one 
non-trivial solution $q_1, q_2 \in \gamma$ 
satisfying 
$q_1+q_2+q_3=0$ 
for a given $q_3 \in \gamma-\{0\}$.
In the rest of this section,
we will show that
if the eight-shaped curve $\gamma$ has some sufficient conditions,
the non-trivial
pair $\{q_1,q_2\}$ 
is unique, smooth and strongly monotonic.
%
The sufficient conditions are the followings;
(V) The curvature of the curve is negative, namely  $f''(x)<0$ for $0<x< 1$.
(VI) The third derivative is also negative,
$f'''(x)<0$ for $0<x< 1$.

Before describe the next theorem,
note that 
the conditions (IV) and (V)
implies that there is a unique value of $x=a_0$ with $0 < a_0 < 1$ that satisfies
\begin{equation}
f'(a_0)=-f'(0).
\end{equation}
We write the point $(a_0, f(a_0))=p_0$.
\begin{theorem}
\label{th:fig8}
If an eight-shaped curve $\gamma$ which is invariant under inversion
$x \mapsto -x$ or $y \mapsto -y$
is described 
in  the first quadrant 
by
a curve $(x, f(x))$ with $0 \le x \le 1$ that satisfies
$f(0)=f(1)=0$, $f'(0)$ is positive finite,
$f'(x) \to -\infty$ for $x \to 1-0$
and for $0 < x < 1$
\begin{eqnarray}
f(x)>0, 
f''(x)< 0,
f'''(x)< 0,
\end{eqnarray}
the solutions of $q_1+q_2+q_3=0$ with $q_1, q_2 \in \gamma$ 
for a given $q_3 \in \gamma - \{0\}$ are two pairs,
trivial one $\{q_1,q_2\}=\{O, -q_3\}$ and non-trivial one $\{q_1,q_2\}=\{q, q^*\}$.
%
For the case $q_3=p_0=(a_0, f(a_0))$,
the trivial pair and the non-trivial pair are coincide,
$\{q_1,q_2\}=\{O, -p_0\}=\{q, q^* \}$ 
where 
$a_0$ 
is the unique solution of $f'(a_0)=-f'(0)$, $0<a_0<1$.
%
When $q_3$ moves around $\gamma$,
the motion $q_i(\sigma)$, $i=1,2$ 
of the non-trivial pair
are smooth, i.e., $|dq_i/d\sigma|< \infty$,
and strongly monotonic i.e., $dq_i/d\sigma\ne 0$ 
where $\sigma$ is the curve length for $q_3$.
\end{theorem}
A proof of this theorem will be given in the following subsections.
%

As mentioned in Section \ref{subsec:convexCurve}, the motion 
of non-trivial pair in this theorem is uniquely parameterized 
as $q_i(\sigma(t))$ by the curve length $\sigma(t)$ of $q_3$.
%
For eight-shaped curves, the area $S$ is zero,
\begin{equation}
S = \frac{1}{2} \oint_{\gamma} q_i \times dq_i=0.
\end{equation}
Therefore the constant angular momentum should be zero,
\begin{equation}
c 
= \frac{1}{T} \int_0^T c\ dt
= \frac{6S}{T}=0,
\label{czero}
\end{equation}
where $T$ is a time in which
the points $q_i$ move around the whole $\gamma$.
Thus, 
unlike Section \ref{sec:convexCurve},
the equation
\[
c
=\frac{d\sigma}{dt}
	\sum_{i=1,2,3} q_i(\sigma)\times \frac{dq_i(\sigma)}{d\sigma}
=0
\]
gives no information
for $d\sigma/dt$.
%
Although
vanishing angular momentum does not determine the speed of the motion,
it imposes a strong constraint on the shape of curve $\gamma$.
Namely, by the three tangents theorem \cite{ffoLem, ffkoy, fujiwaraMontgomery},
three tangent lines at $q_i(\sigma)$ must meet at a point
for each $\sigma$.

Then, we use the energy constant assuming some potential energy $V$,
\begin{equation}
V = \sum_{i<j} U(|q_i-q_j|)
\end{equation}
to determine $\sigma(t)$ by the Hamiltonian $H$,
\begin{equation}
H
=\frac{1}{2}\left( \frac{d\sigma}{dt}\right)^2
		\sum_{i=1,2,3} \left|\frac{dq_i(\sigma)}{d\sigma}\right|^2
		+V
=\textrm{constant}.
\end{equation}
This condition $H=\textrm{constant}$ determines the motion completely,
although
this motion is not guaranteed to satisfy the equation of motion
derived from this Hamiltonian.
%
%
Following the argument in Section \ref{sec:convexCurve}, 
for eight-shaped curves, again
the motion $q_i(\sigma(t))$ are determined completely as 
a ``choreography"
\begin{eqnarray}
q_2(t)&=&q_1(t+T/3), \\
q_3(t)&=&q_1(t+2T/3).
\end{eqnarray}
Thus,
a ``choreography'' in
%
an eight-shaped curve
is determined by the curve $\gamma$ that satisfies the three tangents theorem
that ensures the angular momentum being zero
and by the potential energy.

To prove Theorem \ref{th:fig8}, 
we use Theorem \ref{Ozaki}.
For the curve $\gamma$, we call the right lobe $R$ and left lobe $L$.
Similarly, $R_{\|}$ and $L_{\|}$ for $\gamma_{\|}$.
Note that
$R^*=L_{\|}$ and $L^*=R_{\|}$.
%
In the following subsections,
we give a proof of Theorem \ref{th:fig8} for the cases
$0<a<a_0$, $a=a_0$ and $a_0 < a \le 1$,
where $q_3 = (a, f(a))$,
separately.

\subsection{For the case $0<a<a_0$}
\label{caseBeforeCritical}
\begin{figure}[h] 
   \centering
   \includegraphics[width=8cm]{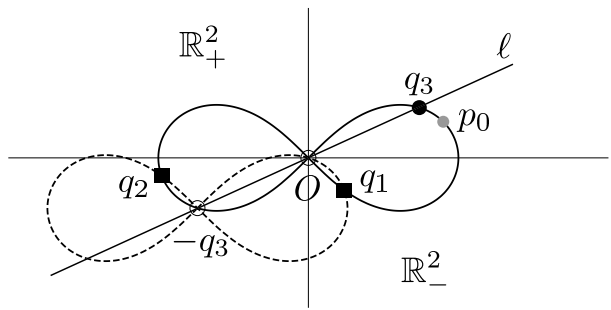} 
   \caption{The curves $\gamma$ (solid line)
   and $\gamma_{\|}$ (dashed line)
   for $0<a<a_0$ where $q_3 = (a,f(a))$.
   Solid black and gray circles represent $q_3$ and $p_0$ respectively. 
   Hollow circles represent the trivial solution $\{O,-q_3\}$
   and solid square represent the non-trivial solution $\{q_1, q_2\}$.
   The line $\ell$ that passes through the points $O$ and $-q_3$
   splits the plane $\mathbb{R}^2$ into
   $\mathbb{R}^2_{+}$, $\mathbb{R}^2_{-}$ and the line $\ell$ itself.
   }
   \label{fig:beforeCritical}
\end{figure}

See Figure \ref{fig:beforeCritical}.

1) For $R \cap R_{\|}$:
It is obvious the origin 
$O \in R \cap R_{\|}$.
Since $f''(x) <0$ and $0<a<a_0$,
we have $f'(0) > f'(a) > f'(a_0)=-f'(0)$.
Therefore $R_{\|}$ starts the origin to inside of $R$.
The lobe $R_{\|}$ cut the $y$-axis at $(0, -2f(a))$ which is obviously
outside of $R$.
Therefore, there is at least one point $q_1 \ne O$ in $R \cap R_{\|}$.
Therefore, $R \cap R_{\|}=\{O,q_1\}$
by \ref{sec:convexcurve} Lemma \ref{lemma:convexcurve}.

2) For $L \cap L_{\|}$:
By the map $q \mapsto q^*$,
we get $L \cap L_{\|}=R_{\|}^* \cap R^*=\{O^*, q_1^*\}=\{-q_3,q_2\}$.

3) For $L \cap R_{\|}=L \cap L^*$:
The line $\ell$ connecting the origin $O$ and $-q_3$
splits the plane $\mathbb{R}^2$ into three parts,
open upper half that we write $\mathbb{R}^{2}_+$,
open lower half  $\mathbb{R}^{2}_-$
and the line $\ell$ itself.
This line also split $L$ and $L^*$ into three parts.
We will show that $L \cap L^* \cap \mathbb{R}^{2}_+$ is empty.
To find the number of elements of $L \cap L^* \cap \mathbb{R}^{2}_+$,
let us consider the difference between 
the $y$ component of the curve $L  \cap \mathbb{R}^{2}_+$
and that of the curve $L^* \cap \mathbb{R}^{2}_+$,
which is described by the following function
\begin{equation}
g(x,a)=f(-x)-f(x+a)+f(a),
\end{equation}
defined in $-a < x <0$.
In \ref{secNumberOfZerosOfG},
we have shown that there is no solution of $g(x,a)=0$
in $-a < x <0$ for $0< a < a_0$.
Therefore, $L \cap L^* \cap \mathbb{R}^{2}_+$ is empty.

By the map $q \mapsto q^*$,
the region $\mathbb{R}^{2}_{+}$ maps onto the region $\mathbb{R}^{2}_{-}$,
therefore, $L \cap L^* \cap \mathbb{R}^{2}_-$ is also empty.
Thus we get
$L \cap L^*=L \cap L^* \cap \ell = \{O, -q_3\}$
since $L$ has at most two common points with any line.

4) For $R \cap L_{\|}$: $R$ is in the region $x \ge 0$, while $L_{\|}=R^*$ is in $ x \le -a<0$.
Therefore, $R \cap L_{\|}$ is empty.

\begin{figure}[h]
   \centering
   \includegraphics[width=8cm]{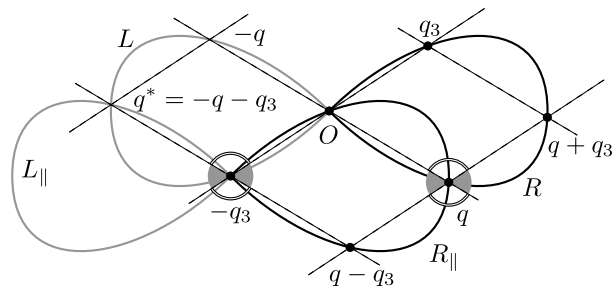} 
   \caption{Lemma \ref{lem:smooth} for eight-shaped curve
   for $0<a<a_0$.
  We denote the cross point of $R$ and $R_\|$ by $q$. 
 (i) The parallelogram $\alpha q + \beta q_3$, $0 \le \alpha, \beta \le 1$ 
   is included inside $R$, 
   thus, at $q$, the tangent line to the curve $\gamma$ passes in the shaded area,
   while
   the tangent line to the curve $\gamma_\|$ at $q$ passes in the non shaded area
   because
   the parallelogram $\alpha q - \beta q_3$, $0 \le \alpha, \beta \le 1$ 
   is included inside $R_{\|}$.
   Therefore,
   the tangent lines at $q$ to the lines $\gamma$ and $\gamma_{\|}$
   are distinct.
  (ii)The parallelogram $-\alpha q - \beta q_3$, $0 \le \alpha, \beta \le 1$ 
   is included inside $L$,
   thus the tangent line to the curve $\gamma$ at $-q_3$ passes in the shaded area.
   Therefore,
   the tangent line to $\gamma_{\|}$ at $q$
   and the tangent line to $\gamma$ at $-q_3$
   are not parallel. }
   \label{fig:smooth_8}
\end{figure}

Summarizing the results of 1)--4), we conclude that
there are 
one trivial 
pair $\{O, -q_3\} \subset \gamma \cap \gamma_{\|}$
and one 
non-trivial 
pair $\{q_1, q_2=q_1^*\} \subset \gamma \cap \gamma_{\|}$
for the case $0<a<a_0$.
%
The smoothness and strong monotonicity of $q_i(\sigma)$ for $i=1,2$
can be proved by Lemma \ref{lem:smooth} referring to Figure \ref{fig:smooth_8}.

\subsection{For the case $a=a_0$}
\label{caseAtCritical}
\begin{figure}[h] 
   \centering
   \includegraphics[width=8cm]{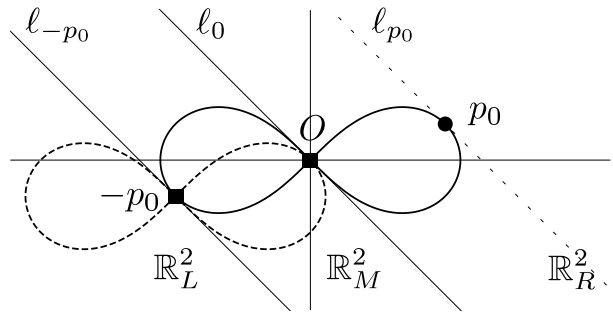} 
   \caption{The case $a=a_0$. The set  $\gamma \cap \gamma_{\|}=\{O, -p_0\}$.
   Two tangent lines $\ell_0$ and $\ell_{-p_0}$ splits the plane $\mathbb{R}^2$
   into open region $\mathbb{R}^2_{L}$, $\mathbb{R}^2_{M}$,$\mathbb{R}^2_{R}$,
  and the lines.
   }
   \label{fig:critical}
\end{figure}

See Figure \ref{fig:critical}.
Let the tangent line of the curve at $p_0=(a_0,f(a_0))$ be $\ell_{p_0}$,
and its parallel translation 
by $q \mapsto q-p_0$ and $q \mapsto q-2p_0$
be $\ell_0$ and $\ell_{-p_0}$.
The lines $\ell_0$ and $\ell_{-p_0}$ are the tangent lines of the curve
$\gamma$ and $\gamma_{\|}$ that pass through the origin and $-p_0$.
The two parallel lines $\ell_0$ and $\ell_{-p_0}$ split 
$\mathbb{R}^2$ into five pieces,
three open $2$-dimensional regions and two lines.
We name the three regions from left to right
$\mathbb{R}^2_{L}$,
$\mathbb{R}^2_{M}$, and 
$\mathbb{R}^2_{R}$.
Obviously, 
$\gamma \cap \gamma_{\|} \cap \mathbb{R}^2_{L}$
and $\gamma \cap \gamma_{\|} \cap \mathbb{R}^2_{R}$
are empty.
%
The set $\gamma \cap \gamma_{\|} \cap \mathbb{R}^2_{M}$
is also empty since, by the same argument 
for $L \cap L^* \cap \mathbb{R}^{2}_+$
in the previous subsection,
$g(x,a_0)=0$ has no solution in $-a<x<0$, which is shown in
\ref{secNumberOfZerosOfG}.

Therefore, 
$\gamma \cap \gamma_{\|}
=\gamma \cap \gamma_{\|} \cap (\ell_{0} \cup \ell_{-p_0})
=\{O,-p_0\}$
for $a=a_0$.

\subsection{For the case $a_0 < a \le 1$}
\label{caseAfterCritical}
\begin{figure}[h] 
   \centering
   \includegraphics[width=8cm]{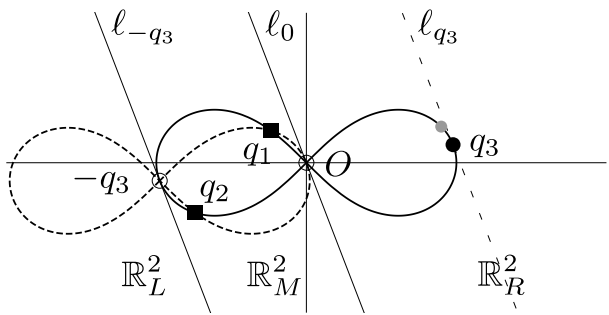} 
   \caption{The case $a_0 < a \le 1$.
   The set  $\gamma \cap \gamma_{\|}=\{O, -q_3\} \cup \{q_1, q_2=q_1^*\}$.
   Three tangent lines $\ell_0$ and $\ell_{-q_3}$ splits the plane $\mathbb{R}^2$
   into open region $\mathbb{R}^2_{L}$, $\mathbb{R}^2_{M}$,$\mathbb{R}^2_{R}$
   and the lines. The non-trivial pair $\{q_1, q_2\}$ is
   in the region $\mathbb{R}^2_{M}$.}
   \label{fig:afterCritical}
\end{figure}

See Figure \ref{fig:afterCritical}.
Similarly to Section \ref{caseAtCritical},
let the tangent line of the curve at $q_3=(a,f(a))$ be $\ell_{q_3}$,
and its parallel translation
by $q \mapsto q-q_3$ and $q \mapsto q-2q_3$
be $\ell_0$ and $\ell_{-q_3}$.
We use the same notations as Section \ref{caseAtCritical},
$\mathbb{R}^2_{L}$,
$\mathbb{R}^2_{M}$, and
$\mathbb{R}^2_{R}$
for the regions. 
Obviously,
$\gamma \cap \gamma_{\|} \cap \mathbb{R}^2_{L}$ and
$\gamma \cap \gamma_{\|} \cap \mathbb{R}^2_{R}$ are empty.
By the same arguments 
for $L \cap L^* \cap \mathbb{R}^{2}_+$
in Section \ref{caseBeforeCritical},
$\gamma \cap \gamma_{\|} \cap \mathbb{R}^2_{M}$
is $\{q_1, q_2=q_1^*\}$
with $q_1=(x_0(a), f(x_0(a)))$ 
where $x=x_0(a)$ is the only solution of $g(x,a)=0$ in $-a<x<0$.
%
See \ref{secNumberOfZerosOfG}.

Therefore, we conclude that the set $\gamma \cap \gamma_{\|}$
has one 
trivial 
pair $\{O, -q_3\}$
and one 
non-trivial
pair $\{q_1, q_2=q_1^*\}$
for the case $a_0 < a \le 1$.
The smoothness and strong monotonicity of $q_i(a)$ for $i=1,2$ 
are explicitly given by the equation (\ref{eq:nonStop})
in \ref{secNumberOfZerosOfG}.

\section{Summary and discussions}
\label{sec:summary}
In this paper,  
we have shown that
the motion of equal mass three bodies in a given curve
is uniquely determined as a choreography 
for the following two cases.
(i) Convex curves that have  point symmetry with respect to 
the origin
and non-vanishing angular momentum are given.
(ii) Eight-shaped curves 
and the energy constant are given.

For eight-shaped curves,
the conditions (V) in Section \ref{sec:Fig8},
the convexity of the each lobe, 
is numerically satisfied by the figure-eight solutions
under homogeneous potential $\alpha^{-1}r^\alpha$ with $\alpha<2$
and proved for the Newtonian potential, $-r^{-1}$, 
by Fujiwara and Montgomery \cite{fujiwaraMontgomery}.
The condition (VI) and all the other conditions
are numerically satisfied by the figure-eight
solution for the Newtonian potential.

Moreover, Theorem \ref{th:fig8} holds 
for the lemniscate curve of Bernoulli although 
it does not satisfy the condition (VI) at a point
 $x_0 = \sqrt{5/32} = 0.395285$, i.e., $f'''(x_0)=0$. 
Therefore we know the only
possible motion of the equal mass three bodies  in the lemniscate are
$(x(\tau(t)), y(\tau(t)))$ with a smooth function $\tau(t)$ 
where 
\begin{eqnarray}
x(t) &=& \frac{\sn(t)}{1+\cn^2(t)}  \nonumber
\\ 
y(t) &=& \frac{\sn(t)\sn(t)}{1+\cn^2(t)},
\end{eqnarray}
and, $\sn$ and $\cn$ are the Jacobian elliptic functions \cite{ffoLem}.

As for the condition (VI), it 
seems too strong
as we have seen in the proof of Theorem \ref{th:fig8}.
Also we note that the all conditions for the theorem are geometric except for 
this condition.
To replace this condition to more weak one
and more geometric quantity 
is a future work.

For a general closed curve,
which is not point symmetric, 
we can investigate the uniqueness of the 
equal mass
three-body 
motion in it in the same manner.
First, we investigate
a non trivial pair $\{q, q^*\}$ for all $q_3 \in \gamma$
in Corollary \ref{cor:OzFkd}.
This might be lengthy and tedious as we did in this paper.
However, once uniqueness and smoothness of the pair was found,
the motion in such curve is determined uniquely modulo
time re-parameterization,
$q_i(t) \mapsto q_i(\sigma(t))$ with function $\sigma(t)$.

To determine the function $\sigma(t)$,
we can use the constancy of the angular momentum,
like Kepler did,
\begin{equation}
\label{eq:angularMomentum}
c
=\sum_{i=1,2,3} q(\sigma(t))\times \frac{dq_i(\sigma(t))}{dt}
= \frac{d\sigma}{dt}
	\sum_{i=1,2,3} q(\sigma)\times \frac{dq_i(\sigma)}{d\sigma}.
\end{equation}
%
The value of the constant angular momentum $c$ is related to
the total area $S$ of the curve. 
%
If the total area $S$ is not zero, then $c\ne 0$.
Then the equation (\ref{eq:angularMomentum}) determines $d\sigma/dt$ and
the motion $q_i(\sigma(t))$ are determined completely.
%
While if $S=0$ like an eight-shaped curve,
then $c=0$.
Thus, the equation (\ref{eq:angularMomentum}) gives no information
for $d\sigma/dt$.
Then, we can use the energy constant assuming some potential energy $V$
to determine $\sigma(t)$ by the Hamiltonian, $H=$ constant.
Thus, 
the motion $q_i(\sigma(t))$ are determined completely.


The angular momentum and the Hamiltonian are invariant
under the exchange of the bodies $1 \to 2 \to 3 \to 1$.
This invariance and the uniqueness of the motion
yield 
the three-body chase
%
with an equal time spacing, namely,
``the choreography in the given curve'',
%
$q_1(t)=q(t)$,
$q_2(t)=q(t+T/3)$,
$q_3(t)=q(t+2T/3)$.

Since the motion $q_i(t)$ is determined uniquely,
the acceleration $d^2 q(t)/dt^2$ is also determined uniquely.
Therefore, whether the equation of motion
\begin{equation}
\frac{d^2 q_i(t)}{dt^2}
=- \frac{\partial V}{\partial q_i}
\end{equation}
with an appropriate potential energy $V$ is satisfied or not
is a test
whether the motion is actually realized by the potential or not.
However, in general, 
it is very hard to find the potential energy $V$ 
which realizes the three-body motion in given curves.

For the figure-eight solution,
the shape of the curve
that corresponds to Kepler's first law
is not known.
The three tangents theorem \cite{ffoLem,ffkoy,fujiwaraMontgomery}
is  a strict constraint for the curve
and would be a clue to find it.

Finally, 
one may consider the general three-body problem 
in given curves,
where the masses and orbits of three bodies
are not equal,
%
using
Theorem \ref{Ozaki-Fukuda}.

\appendix

\section{Number of cross points of convex curve $\gamma$ and $\gamma_{\|}$}
\label{sec:convexcurve}
\begin{lemma}
\label{lemma:convexcurve}
Consider a closed convex curve $\gamma$ in $\mathbb{R}^2$
that has at most two common points with any line.
Then, the cross points of $\gamma$ and 
its parallel translation
$\gamma_{\|}=\{ q+p | q \in \gamma \}$
with $p\ne 0$
are at most two.
\end{lemma}

\noindent
\textbf{Proof:}
Suppose there are three distinct points $a_1$, $a_2$ and $a_3$
$\in \gamma \cap \gamma_{\|}$.
Then, by the definition, points 
$a'_1=a_1+p$, $a'_2=a_2+p$ and $a'_3=a_3+p$ are also
in $\gamma$.
Therefore, the points $a_1, a_2, a_3, a'_1, a'_2, a'_3$ are in $\gamma$.

Take a coordinate system whose $x$-axis is parallel to the line $a'_i a_i$
so that the $x$ components of $a'_i$ are larger than that of $a_i$,
namely, $x'_i = x_i+|p|$.
Then points $a_i$ and $a'_i$ have the same $y$ component $y_i$.
Rename the points so that the $y$ component of the points are
$y_1 < y_2 < y_3$. (These values are distinct, otherwise more than three points
are in a line.)
Take a oblique coordinates whose $y$-axis is parallel to the line $a_1 a_3$.
Then the components of the points are
$a_1=(x,y_1)$,
$a_2=(x_2,y_2)$,
$a_3=(x,y_3)$,
$a'_1=(x+|p|, y_1)$,
$a'_2=(x_2+|p|, y_2)$, 
$a'_3=(x+|p|, y_3)$.

\begin{figure}[h] 
   \centering
   \includegraphics[width=6cm]{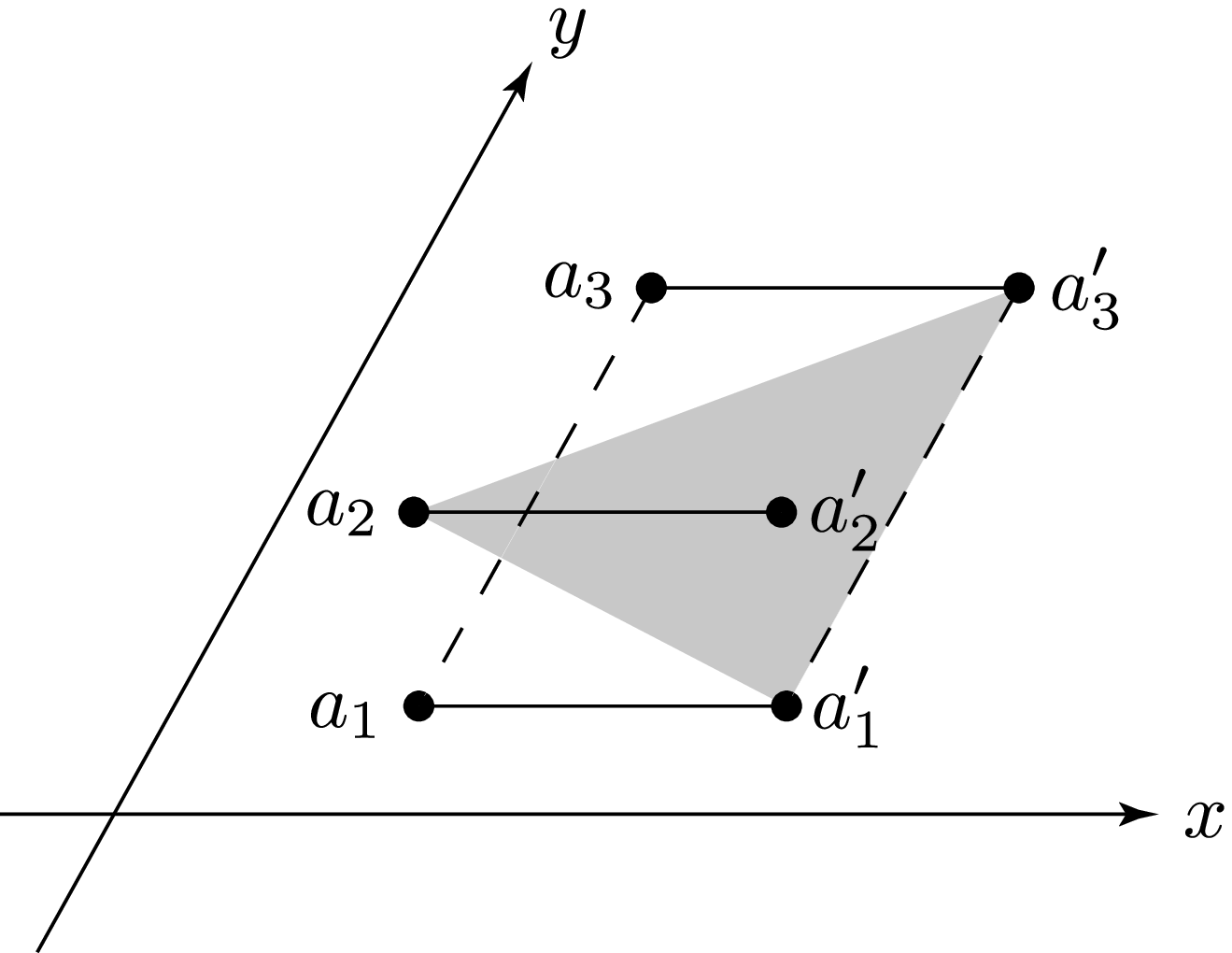} 
   \caption{If the $X$-coordinate of $a_2$ is smaller than
   that of $a_1$ and $a_3$, then the point $a'_2$
   is inside of the triangle $a'_1 a_2 a'_3$.}
   \label{fig:figPositionOfa2}
\end{figure}

If $x_2 < x$ then $a'_2$ is in the triangle $a'_1 a_2 a'_3$.
See Figure \ref{fig:figPositionOfa2}.
If $x < x_2$ then $a_2$ is in the triangle $a_1 a'_2 a_3$.
Both cases contradict to the convexity of the curve $\gamma$.
If $x_2=x$ then three points $a_1$, $a_2$ and $a_3$ 
are in a line,
which contradicts to the assumption of the lemma. 

This contradiction comes from the assumption of the existence
of the three distinct points in $\gamma \cap \gamma_{\|}$.
Thus, we prove this lemma.

Note that if the curvature of a closed convex curve 
is not zero almost everywhere, 
the curve has at most two common points with any line.
Therefore the closed convex curve in Theorem \ref{th:convex}, and 
the lobe $R$ or $L$ of the eight-shaped curve in Theorem \ref{th:fig8}
satisfy the conditions for this Lemma.

\section{Number of zeros for a function}
\label{secNumberOfZerosOfG}
Let $f(x)$ be a function defined in the region $0 \le x \le 1$
and satisfies the following properties,
\begin{eqnarray}
f(0)=f(1)=0,\\
f(x) > 0, f''(x)<0, f'''(x)<0 \mbox{ for } 0<x<1.
\end{eqnarray}
Thus
there exists a unique value $a_0$ in $0< a_0 < 1$ such that
\begin{equation}
f'(a_0)=-f'(0)<0.
\end{equation}

We define the following function $g(x,a)$,
\begin{equation}
g(x,a) = f(-x) - f(x+a)+f(a),
\end{equation}
in the region $0<a \le 1$, $-a \le x \le 0$.
For a while,
we consider the behavior of $g(x,a)$ for fixed value of $a$.
So, we simply write
\begin{equation}
g(x)=g(x,a) \mbox{ for fixed }a
\end{equation}
and 
\begin{equation}
g'(x) = \frac{\partial g(x,a)}{\partial x}.
\end{equation}

We will show that the number of solutions of 
$g(x)=0$ in $ -a < x < 0$ 
is zero for $0 < a \le a_0$
and one for $a_0 < a \le 1$.

Since,
\begin{equation}
g'''(x)=-f'''(-x)-f'''(x+a)>0
\end{equation}
and
\begin{equation}
g''(-a/2)
=f''(a/2)-f''(a/2)
=0,
\end{equation}
we get
\begin{eqnarray}
g''(x) < 0 \mbox{ for } -a \le x < -a/2,\label{eq:g''1}\\
g''(x) > 0 \mbox{ for } -a/2 < x \le 0.\label{eq:g''2}
\end{eqnarray}
Note that
\begin{eqnarray}
&&g(-a) = 2f(a) \ge 0,\\
&&g(-a/2) = f(a) \ge 0,\\
&&g(0)=0,
\end{eqnarray}
and
\begin{equation}
g'(0)=-f'(0)-f'(a)=f'(a_0)-f'(a)
\end{equation}
is a increasing function of $a$
because $f''(x)<0$.

\begin{figure}[htbp] 
   \centering
   \includegraphics[width=15cm]{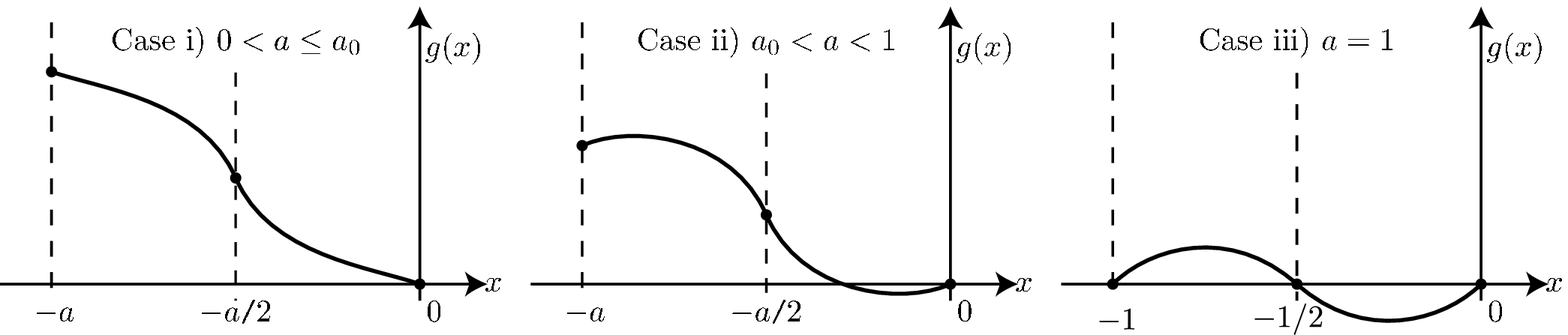}
   \caption{Schematic view of $g(x)=f(-x)-f(x+a)+f(a)$.
The curve $(x,g(x))$ is point symmetric with respect to $\left(-a/2, g(-a/2)\right)$, namely,
$g(-x-a/2)-g(-a/2)=- \left( g(x-a/2)-g(-a/2)\right)$.

}
   \label{figGforCasesItoIII}
\end{figure}

We split the problem into three cases by the value of $a$.
See Figure \ref{figGforCasesItoIII}.

i) The case $0 < a \le a_0$:
We have $g'(0)=f'(a_0)-f'(a)\le 0$.
Because $f(a)>0$, we have $g(-a)>0$, $g(-a/2)>0$ and $g(0)=0$.
Then inequalities (\ref{eq:g''1}) and (\ref{eq:g''2})
prove that there is no solution of $g(x)=0$ in $-a < x <0$ in this case.

For the case $a=a_0$,
we have $g(0)=g'(0)=0$ and $g''(0)>0$.
Therefore, $g(0)=0$ is a double root.

ii) The case $a_0 < a <1$:
We have $g'(0)>0$, $g(-a)>0$, $g(-a/2)>0$ and $g(0)=0$.
Therefore, inequalities (\ref{eq:g''1}) and (\ref{eq:g''2}) prove that
there is no solution of $g(x)=0$
in $-a < x \le -a/2$
and one solution in $-a/2 < x <0$.
Let us write the zero point $x_0(a)$.
Note that $g'(x)$ at $x=x_0(a)$ is negative, namely
\begin{equation}
\label{eq:negative}
g'(x_0(a))=-f'(-x_0(a))-f'(x_0(a)+a)<0.
\end{equation} 

iii) The case $a=1$:
We have $g(-a)=g(-a/2)=g(0)=0$.
Therefore, the equations (\ref{eq:g''1}) and (\ref{eq:g''2}) prove
that the $x=-a/2$ is the only solution of $g(x)=0$ in $-a<x<0$.
The inequality (\ref{eq:negative}) is also true for this case.

Now, let us consider the behavior of the zero point of $g$ for the range $a_0<a\le 1$.
To do this, it is convenient to use the full expression $g=g(x,a)$.
%
%
Then we have
\begin{equation}
g(x_0(a),a)=f(-x_0(a))-f(x_0(a)+a)+f(a)=0
\end{equation}
for all $a_0<a\le 1$.
Then, total derivative of this expression by $a$ yields
\begin{equation}
0
=\frac{dg(x_0(a),a)}{da}
=\left. \frac{dx_0(a)}{da}\frac{\partial g(x,a)}{\partial x} + \frac{\partial g(x,a)}{\partial a}
	\right|_{x=x_0(a)}.
\end{equation}
\begin{equation}
\therefore
\frac{dx_0(a)}{da}\Big(f'(-x_0(a))+f'(x_0(a)+a)\Big)
	=f'(a)-f'(x_0(a)+a).
\end{equation}
By (\ref{eq:negative}), we have $f'(-x_0(a))+f'(x_0(a)+a)>0$.
While $f'(a)-f'(x_0(a)+a)<0$ by $f''(x)<0$.
Thus, we get
\begin{equation}
\label{eq:nonStop}
\frac{d x_0(a)}{da}<0 \mbox{ for } a_0 < a \le 1,
\end{equation}
namely,
the zero point of $g(x)=0$ appears near the origin and moves to $-1/2$
smoothly and strongly monotonically when $a$ increases $a_0$ to $1$.

\ack
Toshiaki Fujiwara was supported by
Grant-in-Aid for Scientific Research (C)
19540227.



\Bibliography{9}

\bibitem{moore}
Moore C 1993
Braids in Classical Gravity
{\it Phys. Rev. Lett.} {\bf 70} 3675--3679

\bibitem{chenAndMont}
Chenciner A and Montgomery R 2000
A remarkable periodic solution of the three-body problem in the case of equal masses
{\it Annals of Mathematics} {\bf 152} 881--901

\bibitem{simo1}
Sim\'{o} C 2001
Periodic orbits of planar N-body problem with equal masses 
and all bodies on the same path
{\it Proceed. 3rd European Cong. of Math., Progress in Math.}
{\bf 201} (Birk\"auser, Basel) 101--115

\bibitem{simo2}
Sim\'{o} C 2002 
Dynamical properties of the figure eight solution of the three-body problem
{\it Celestial mechanics: Dedicated to Donald Saari for his 60th Birthday.
Contemporary Mathematics} {\bf 292}
(Providence, R.I.: American Mathematical Society)
209--228

\bibitem{fujiwaraMontgomery}
Fujiwara T and Montgomery R 2005
Convexity of the figure eight solution to the three-body problem.
{\it Pacific Journal of Mathematics} {\bf 219} 271--283



\bibitem{ffoLem}
Fujiwara T, Fukuda H and Ozaki H 2003
Choreographic three bodies on the lemniscate
{\it J. Phys. A: Math. Gen.}
{\bf 36}
2791--2800

\bibitem{ffkoy}
Fujiwara~T, Fukuda~H, Kameyama~A, Ozaki~H and Yamada~M 2004
Synchronized similar triangles for three-body orbits with zero angular momentum,
{\it J. Phys. A: Math. Gen.}
{\bf 37}
10571--10584








%



\endbib
\end{document}